\RequirePackage{lineno}
\documentclass[aps,amsmath,showkeys,superscriptaddress]{revtex4}        
\usepackage{graphicx}
\usepackage{bbold}
\usepackage{amssymb}
\usepackage[latin1]{inputenc}
\usepackage{xspace}
\usepackage{graphics}
\usepackage{setspace}
\usepackage{tabularx}
\usepackage{threeparttable} 
\usepackage{dcolumn,amsmath}
\usepackage{lscape}
\usepackage{mathrsfs}
\usepackage{rotating}
\usepackage{amsfonts}
\usepackage{amsthm}
\usepackage{bm}
\usepackage{textcomp}
\usepackage{subfig}

\usepackage{lineno}
\usepackage{booktabs}
\usepackage{lmodern}
\usepackage{float}
\usepackage{xcolor}
\usepackage{wasysym} 

\usepackage{multirow}
\usepackage{natbib}
\begin{document}

\title{\linespread{1.5}\selectfont Immunocto: a massive immune cell database auto-generated for histopathology}

\linespread{1.5}
\author{Mika\"el Simard}\email{m.simard@ucl.ac.uk}
\affiliation{Department of Medical Physics and Biomedical Engineering, University College London, London, UK}

\author{Zhuoyan Shen}
\affiliation{Department of Medical Physics and Biomedical Engineering, University College London, London, UK}

\author{Konstantin Br\"autigam}
\affiliation{Centre for Evolution and Cancer, The Institute of Cancer Research, London, UK}

\author{Rasha Abu-Eid}
\affiliation{College of Medicine and Health, University of Birmingham, Birmingham, UK}

\author{Maria A. Hawkins}
\affiliation{Department of Medical Physics and Biomedical Engineering, University College London, London, UK}
\affiliation{Department of Radiotherapy, University College London Hospitals, London, UK}

\author{Charles-Antoine Collins Fekete}
\affiliation{Department of Medical Physics and Biomedical Engineering, University College London, London, UK}

\begin{abstract} 
With the advent of novel cancer treatment options such as immunotherapy, studying the tumour immune micro-environment (TIME) is crucial to inform on prognosis and understand potential response to therapeutic agents. A key approach to characterising the TIME may be through combining (1) digitised microscopic high-resolution optical images of hematoxylin and eosin (H\&E) stained tissue sections obtained in routine histopathology examinations with (2) automated immune cell detection and classification methods. In this work, we introduce a workflow to automatically generate robust single cell contours and labels from dually stained tissue sections with H\&E and multiplexed immunofluorescence (IF) markers. The approach harnesses the Segment Anything Model and requires minimal human intervention compared to existing single cell databases. With this methodology, we create Immunocto, a massive, multi-million automatically generated database of 6,848,454 human cells and objects, including 2,282,818 immune cells distributed across 4 subtypes: CD4$^+$ T cell lymphocytes, CD8$^+$ T cell lymphocytes, CD20$^+$ B cell lymphocytes, and CD68$^+$/CD163$^+$ macrophages. For each cell, we provide a $64\times64$ pixels$^2$ H\&E image at $\mathbf{40}\times$ magnification, along with a binary mask of the nucleus and a label. The database, which is made publicly available, can be used to train models to study the TIME on routine H\&E slides. We show that deep learning models trained on Immunocto result in state-of-the-art performance for lymphocyte detection. The approach demonstrates the benefits of using matched H\&E and IF data to generate robust databases for computational pathology applications.


\noindent ---------------------------------------------------------------------------------------------------------------------------------

\end{abstract}

\keywords{computational pathology, immunotherapy, lymphocytes, macrophages, segment anything, database}

\maketitle

\linespread{1.25}\selectfont


\section{Introduction}

\label{sec:overview}
\noindent Histopathology relates to the observation of abnormal tissue at the microscopic scale. The gold standard methodology for histopathology is the study of formalin-fixed paraffin-embedded (FFPE) fixed tissue sections under a microscope stained with haematoxylin and eosin (H\&E). To identify diseases, pattern recognition at various scales on tissue samples is performed by an histopathologist. The advent of digital pathology, where tissue sections are digitised to create gigapixel whole slide images (WSIs), opens the possibility of creating large databases to train models that may aid pathologists in their study of diseases and decision making processes. \\

\noindent In the study of cancer, understanding the role of the tumour immune micro-environment (TIME) is essential \cite{binnewies2018understanding}, as it can advise on prognosis, guide treatment \cite{fridman2012immune,chew2012immune}, or be examined retrospectively to understand response to therapies. For instance, an exploration of the TIME may reveal biomarkers that could be useful in identifying patient populations that respond well to immunotherapy \cite{binnewies2018understanding}. \\

\noindent Microscopic high-resolution imaging, where immune cell location and function can be identified, is a critical next step to understanding the TIME, as it can provide information on cellular proportions, heterogeneity and spatial distribution \cite{binnewies2018understanding}. However, manual annotation of immune cells by pathologists is time-consuming and identifying various immune cell subtypes can be tedious or even infeasible on H\&E WSIs \cite{graham2021lizard}. As such, there is growing interest in developing models that can automatically segment and classify immune cells on WSIs \cite{graham2019hover}. \\

\noindent In this work, we report the creation of a novel database to (1) support the development of new immune cell classification models, and (2) provide a benchmark for evaluating the performance of existing models. While there are existing databases of individually annotated immune cells (table see \ref{tab:datasets} and section \ref{sect:related_work}), they come with their own limitations, which intrinsically constrains the quality of immune cell detection models. First, existing datasets involve important human intervention to generate cell labels and/or nuclei masks, which limits the number of labelled immune cells that can be produced. Second, as existing datasets are obtained using H\&E data only, the accuracy of the underlying labels is limited. For instance, the Lizard database \cite{graham2021lizard} has reported limited inter-observer agreement for immune cell identification. \\

\noindent To solve the above problems, we propose a novel data processing pipeline to automatically generate a massive database of robust single immune cell annotations. The method is based on co-localised, dually stained slides (H\&E and multiplexed immunofluorescence (IF)) to facilitate cell labeling. IF is a technique that uses antibodies conjugated with fluorophores to detect specific proteins on or within cells. When incubated with tissue sections, the fluorescent antibodies will selectively bind to specific antigens. Irradiating the sample using a laser with a wavelength that matches a fluorophore's absorption spectrum will create an IF image of the tissue specimen, revealing locations on the tissue where a particular antigen is expressed. For instance, the presence of CD4$^+$ lymphocytes can be revealed with CD4 antibodies. In that context, IF can be used to characterise various cell types based on their unique surface markers, including CD4$^+$, CD8$^+$ T cells, or CD20$^+$ B cells. \\

\noindent First, the methodology uses the Segment Anything Model (SAM) \cite{kirillov2023segment} on H\&E data to automatically generate high-quality object masks at the cellular level. This includes nuclear structures, epithelial and endothelial cells, muscular and fat cells, apoptotic bodies, blood cells, immune cells, and a variety of tissue artifacts. To facilitate cell labeling, we use the co-localised multiplexed IF signal within each object to identify and subtype immune cells. \\

\noindent We apply this methodology to the Orion dataset \cite{lin2023high} (distributed under the MIT license), which contains H\&E images from colorectal cancer patients co-localised with 18 IF data channels, to create Immunocto, a high-resolution ($\mathbf{40}\times$ magnification, pixel size of 0.325 $\mu$m) massive database of 2,282,818 immune cells distributed across 4 immune cell subtypes (CD4$^+$ T cells, CD8$^+$ T cells, CD20$^+$ B cells, and CD68$^+$/CD163$^+$ macrophages). To our knowledge, Immunocto is the largest available dataset of immune cells extracted from H\&E WSIs by an order of magnitude (table \ref{tab:datasets}). \\

\noindent In this paper, we detail the methodology used to create Immunocto, evaluate the consistency of the database and evaluate the performance of H\&E classifiers trained on Immunocto against other existing immune cell classifiers. The main contributions of this work are summarised as follows:

\begin{itemize}
\item A general methodology to produce robust single cell labels using joint H\&E and multiplexed IF data with minimal annotations. The method can easily be expanded to other cell types. 
\item Immunocto, a new openly available dataset of 2,282,182 immune cell distributed across 4 immune cell subtypes obtained from 40 colorectal cancer patients. The data is accessible at \url{https://zenodo.org/records/11073373}.
\item A demonstration that deep learning models trained on Immunocto achieved state-of-the-art performance for immune cell detection on external H\&E datasets.
\item Cell classification networks that can produce previously unachievable subtyping of immune cells (CD4$^+$ T cells, CD8$^+$ T cells, and CD20$^+$ B cells).
\end{itemize}

\section{Related work}
\label{sect:related_work}

\noindent Considerable efforts have been made to build labeled data sets for nuclear segmentation and/or classification on H\&E WSIs. The MoNuSAC dataset \cite{verma2021monusac2020} contains over 25,000 manually labelled immune cells. Manual labelling of various type of immune cells, which often present similar morphological features, can be cumbersome, requires extensive pathologist time and does not scale well. The PanNuke database \cite{gamper2020pannuke} rather relied on active learning, where pre-labelled data is used to train a cell subtype classifier, and the results are manually verified by pathologists. The PanNuke database is obtained after 7 iterations between model training and pathologist verification. Still, manual investigation has a limited inter-observer reproducibility due to the above-mentioned reason, which limits the accuracy of this methodology.  Alternatively, the larger NuCLS database \cite{amgad2022nucls} was obtained with a crowdsourcing approach combining pathologists and non-pathologists (medical students and graduates), which also needs important human intervention and is still limited in accuracy by human identification on H\&E images only. More recently, the Lizard database \cite{graham2021lizard} uses a four-step approach: iterative segmentation and classification, cell boundary refinement, inflammatory nuclei subtyping, and manual class refinement. Each step involves important pathologist annotation, and the overall process includes the manual annotation of over 70,000 immune cells. \\

\noindent Furthermore, there have only been few investigations into subtyping lymphocytes, which is a difficult and subjective task on H\&E images due to morphological similarities between lymphocyte types \cite{alexander1975human, strokotov2009there}. For instance, lymphocytes can be, amongst others, classified as CD8$^+$ (cytotoxic) T cells, CD4$^+$ T-helper cells, or B cells. Each type has a different function, and capturing the spatial distribution of lymphocytes is an important part of understanding the TIME \cite{binnewies2018understanding}. As an example, stratification of colorectal cancer patients has been correlated with deep infiltration of CD8$^+$ T cells \cite{mlecnik2016integrative}, and the CD4$^+$/CD8$^+$ ratio of tumour-infiltrating lymphocytes has high prognostic value in triple-negative breast cancer \cite{wang2017cd4}. 

\begin{table}[h!]
\caption{Fully labeled and segmented immune cells datasets ($\geq$ 10000 cells) on H\&E WSIs. For cell subtypes, L is lymphocyte, M is macrophage, N is neutrophil, P is plasma cell, and E is eosinophil.}
\label{tab:datasets}
\centering
\begin{tabular}{cccc}
\hline\hline
Dataset & Cells & Cancer location & Cell subtypes \\
\cline{1-4}
MoNuSAC \cite{verma2021monusac2020} & 25,157 & \small Breast, Kidney, Lung, Prostate & L, M, N \\
PanNuke \cite{gamper2020pannuke} & 32,276 & 19 sites & L, M, P \\
NuCLS \cite{amgad2022nucls} & 47,113 & Breast & E, L, M, N, P \\
Lizard \cite{graham2021lizard} & 138,307 & 19 sites & E, L, N, P  \\
\textbf{Immunocto} & \textbf{2,282,818} & Colon & L (\scriptsize CD4$^+$ T cell, CD8$^+$ T cell, B cell)\normalsize, M \\
\hline\hline
\end{tabular}
\end{table}

\section{Methods}

\subsection{Creating Immunocto}
\label{sect:create_immunocto}

\begin{figure}[h!]
\centering
{\includegraphics[width=.83\textwidth]{{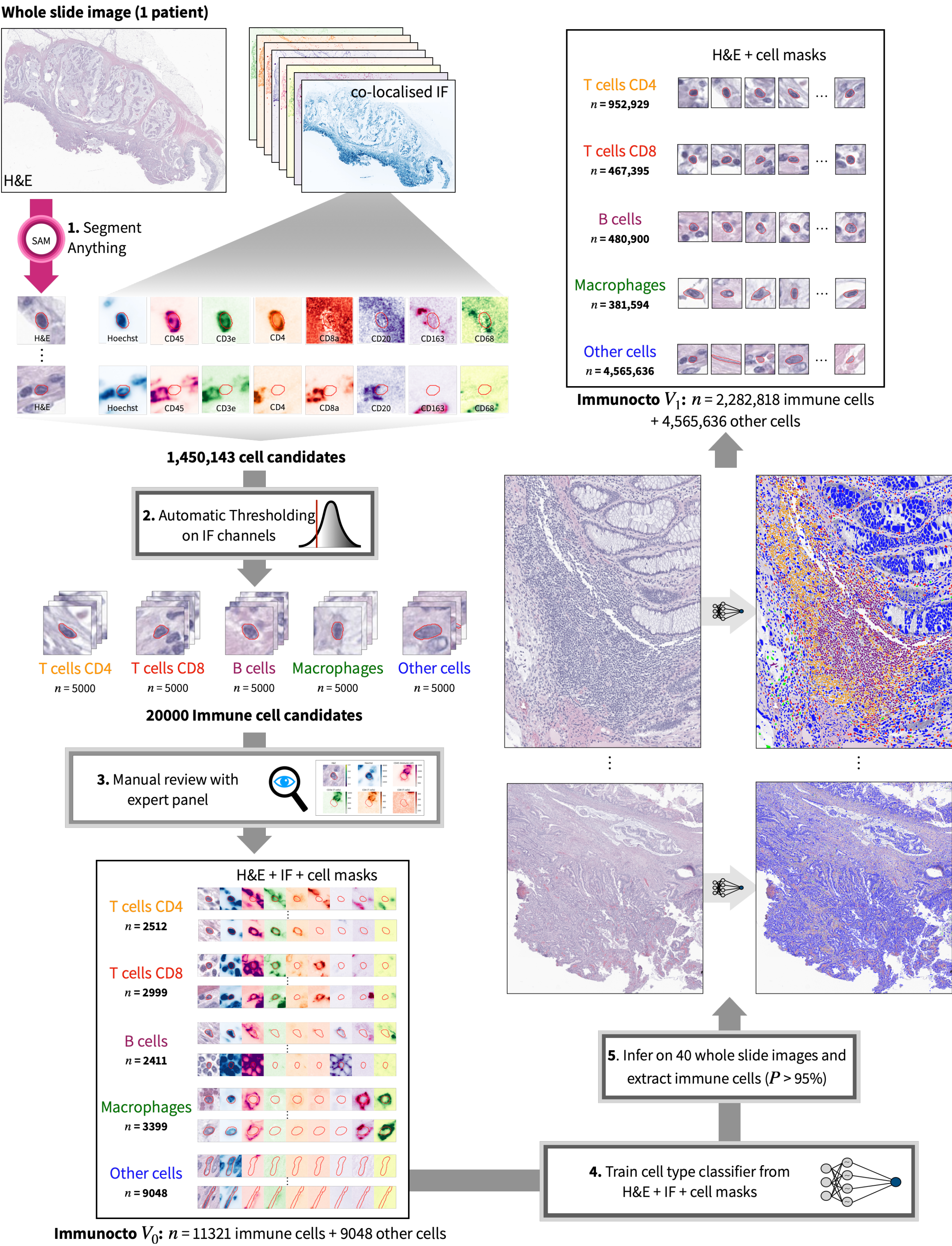}} 
\caption{Workflow to create the Immunocto database.}
\label{fig:immunocto}
}
\end{figure}

\noindent The overall framework for creating Immunocto is illustrated in figure \ref{fig:immunocto}; the 5 steps shown in the figure are detailed in sections \ref{sect:step1_sam} to \ref{sect:step5_immunoctov1}. Briefly, SAM is used on a single WSI to detect objects, which are mostly cell nuclei. Using co-localised IF data and a thresholding approach, 20,000 candidate immune cells are automatically extracted from the SAM-derived masks. The cells are manually reviewed by non-pathologists by looking at relevant IF channels. The result is Immunocto $V_0$, the first iteration of the database containing 11,321 immune cells and 9048 other cells. This database is used to train a model which predicts an immune cell type using H\&E, IF images, and the binary mask from SAM. The model is ran on 40 WSIs with matched H\&E and IF data, and a final, highly specific database, Immunocto $V_1$, is extracted by only preserving cells with high classification probability ($\geq$ 95\%).

\newpage
\subsubsection{Raw data source}
\label{rawdatasource}

\noindent To create Immunocto, we use 40 colorectal cancer patients with publically available, registered H\&E and IF data from the Orion platform proposed by Lin et al. \cite{lin2023high}. WSIs are obtained at $\mathbf{40}\times$ magnification with a pixel size of 0.325 $\mu$m. IF data refers to a set of 18 fluorescence images which are registered to H\&E WSIs and obtained via incubation with various IF markers. Briefly, a tissue section is incubated simultaneously with 18 IF markers, each tagging a specific category of cell or structure. For instance, the Hoechst marker tags DNA, which identifies cell nuclei; the CD45 (leukocyte common antigen) marker tags white blood cells in general, and CD3 tags T cells. A fluorescence microscopy system using various excitation lasers and tunable filters captures an image for each IF marker, and the pixel-wise intensity of each image is proportional to the concentration of the dye in that pixel. H\&E staining and imaging is then performed on the same tissue section. Examples of H\&E along with IF data can be seen in figure \ref{fig:immunocto}.\\

\noindent To extract specific immune cells from IF channels, Lin et al. \cite{lin2023high} proposed a panel for colorectal cancer, i.e. a decision tree where each node considers the presence of a specific IF marker signal. This allows one to identify cell types based on a signature of IF activity. In figure 1 of supplementary material, we present an adapted panel to identify cell types present in Immunocto. With thresholding operations on selected IF channels, one can extract regions of interest (ROIs) associated with specific cell types and generate ground truth labels - this is detailed in section \ref{sect:step2_thresh}.


\subsubsection{Step \#1: detecting cells with Segment Anything}
\label{sect:step1_sam}

\noindent Cell nuclei in a given WSI can be accurately segmented using SAM \cite{kirillov2023segment} (ViT-H architecture). To extract nuclei masks, a WSI is tessellated into non-overlapping tiles of 512$\times$512 pixels$^2$. Each tile is resampled to 1024$\times$1024 pixels$^2$, and a grid of 64$\times$64 uniformly distributed points across the image is used as a point prompt. Resulting masks are filtered with the following empirically defined criteria: predicted intersection over union (IoU) score of $\geq$ 0.81, a stability score $\geq$ 0.85, and an area between 42 and 4320 pixels$^2$ to cover typical immune cell sizes. Masks with overlapping bounding boxes of IoU $>$ 0.1 are discarded via non-maximum suppression using the \verb+batched_nms+ operator from PyTorch, with scores interpreted as the original IoU provided by SAM. Resulting masks are arrays of 64$\times$64 pixels$^2$. An example of cell detection with SAM over a large ROI is shown in figure \ref{fig:side2side}a. Using SAM generally identifies single cells in WSIs, although other objects such as groups of cells, gaps and holes can be also captured.

\subsubsection{Step \#2: Generating immune cell candidates via automatic thresholding.}
\label{sect:step2_thresh}

\begin{figure}[h!]
\centering
{\includegraphics[width=.7\textwidth]{{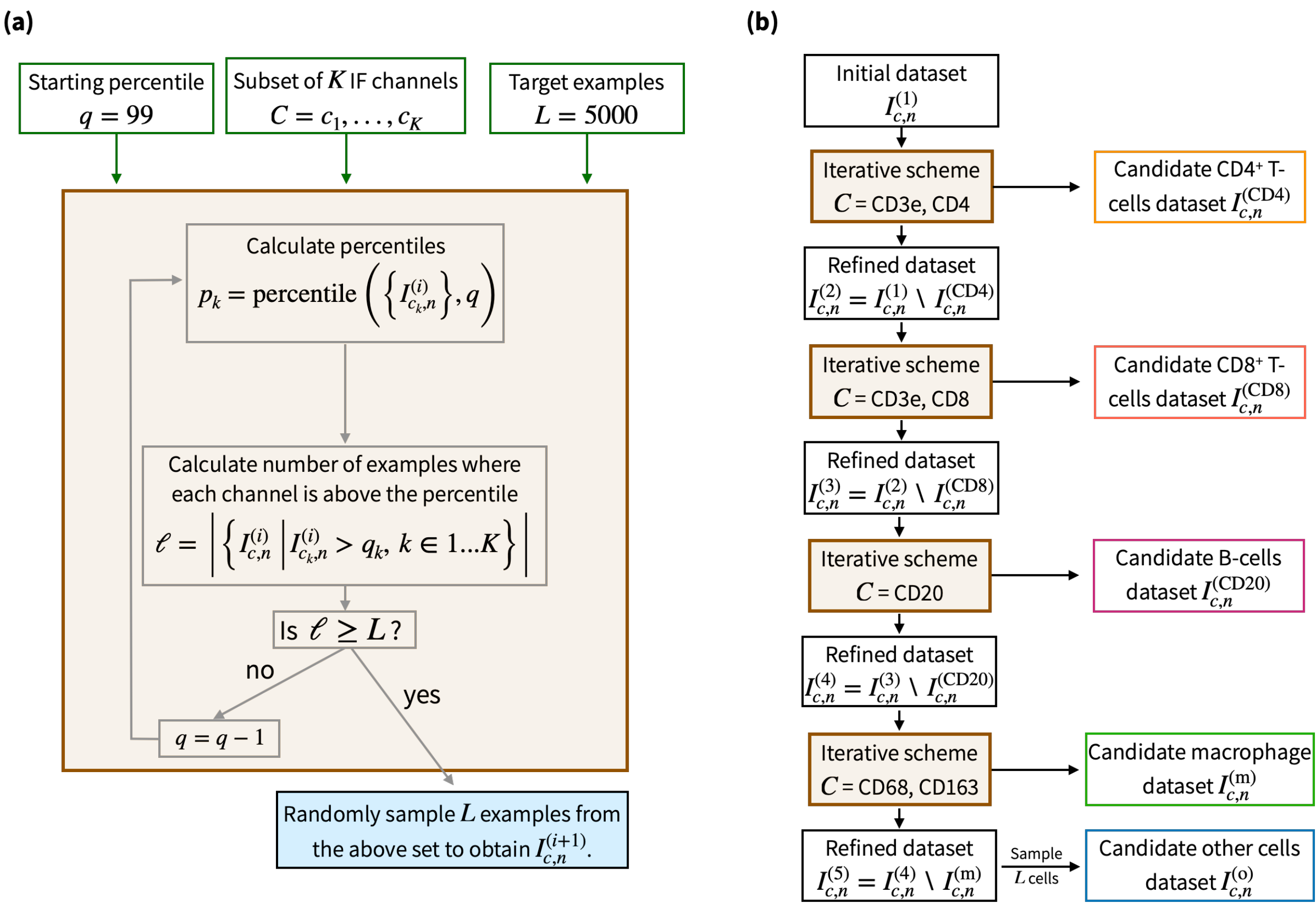}} 
\caption{\textbf{(a)} Iterative thresholding scheme. Given starting parameters $q$, $C$ and $L$, one can extract $L$ candidate cells that are potentially active in immunofluorescence channels $c_1, ..., c_K$. \textbf{(b)} Successive applications of the iterative scheme to extract candidate immune cell datasets. All applications of the iterative scheme use $L=5000$ and $q=99$.}
\label{fig:itsch}
}
\end{figure}

\noindent Applying SAM on a WSI results, on average, in $>$1,000,000 detected objects, most of which are cell nuclei. Leveraging IF data, one can extract candidate objects for immune cells using an intensity thresholding scheme detailed hereafter. Let $\mathcal{I}_{c,n}$ represent the average intensity of IF channel $c$ inside the mask of the $n^{th}$ object detected by SAM in a given ensemble of WSIs. The application of SAM to an ensemble of WSIs results in dataset $\left\{\mathcal{I}^{(0)}_{c,n}\right\}$. Objects which are likely to be cell nuclei are extracted from $\left\{\mathcal{I}^{(0)}_{c,n}\right\}$ using a thresholding operation on the Hoechst channel, an indicator of DNA. This produces $\left\{\mathcal{I}^{(1)}_{c,n}\right\}$, a subset of $\left\{\mathcal{I}^{(0)}_{c,n}\right\}$:

\vspace{-0.5cm}
\begin{equation}
\mathcal{I}^{(1)}_{c,n} = \left\{\mathcal{I}^{(0)}_{c,n} ~\left|~ \mathcal{I}_{\mathrm{Hoechst},n}^{(0)} > \mathcal{H}\right.\right\},
\label{eq:thresh}
\end{equation} 

\noindent where $\mathcal{H}$ is an intensity thresholding value on the Hoechst channel of the WSI, which is arbitrarily set to 30\% of the value obtained with Otsu's method \cite{ostu1979threshold}. The 30\% fraction is selected to compensate for image registration errors between H\&E and the IF channels. To extract immune cell candidates from $\left\{\mathcal{I}^{(1)}_{c,n}\right\}$, the iterative thresholding scheme presented in figure \ref{fig:itsch}a is used. After defining a starting percentile value $q$, a target number of examples $L$ and a subset of $K$ immunofluorescence channels $C=c_1, ..., c_K$ on which to operate, the method iteratively finds the top $L$ examples in the dataset which have maximum signal intensity for all channels. \\

\noindent We combine this iterative thresholding scheme with the IF decision tree (figure 1, supplementary material) to create a workflow that automatically extracts candidates for immune cells (figure \ref{fig:itsch}b). We apply this workflow on the first patient of the Orion cohort (CRC01) to generate 25,000 candidate objects. This includes 5000 cells for each of the four immune cell type - CD4$^+$ T cell lymphocytes, CD8$^+$ T cell lymphocytes, B cells and macrophages, as well as 5000 other objects. While the other objects are mostly non-immune cells, they can also be other objects such as gaps and holes, as equation \ref{eq:thresh} is not a perfect classifier to separate cells from non-cell objects.

\subsubsection{Step \#3: Manual reviewing of the initial database}

\noindent The automatic thresholding method produces misclassified candidate cells due to imperfections in IF staining. This includes slight registration errors between IF and H\&E channels, non-specific targeting of IF markers and a mismatch between the SAM masks, which contour the nuclei, and the IF signal, typically located in the cytoplasm of cells around the nuclei. Therefore, all 25,000 candidate cells are manually reviewed using the IF decision tree (figure 1, supplementary material). The presence of the IF channels allows to confidently classify cells, as it is not based on morphological features of H\&E images but mainly on the intensity of selected IF channels. Manual review was performed by three authors of this study (MS, ZS, CACF) using Label Studio v1.10.1 (\url{https://labelstud.io}). Some cells marked as uncertain by one individual were reviewed by at least another person before assigning a label. \\

\noindent Candidate immune cells were rejected if their IF marker signal was faint and too close to the background noise level. In some limited cases, the masks extracted from SAM included more than a single cell, and those instances were rejected. Through this process, we rejected 4,631 cells out of the initial 25,000 which were not clearly identifiable as single cells corresponding to one class. The resulting database, Immunocto $V_0$, contains 20,369 cells distributed over 5 classes: 11,321 immune cells (2512 CD4$^+$ T cells, 2999 CD8$^+$ T cells, 2411 B cells and 3399 macrophages), as well as 9048 other cells.

\subsubsection{Step \#4: Training a immune cell classifier from H\&E, IF and cell masks}
\label{sect:step4_trainclassifonv0}

\noindent The above step \#3 - recognising immune cells based on matched IF and H\&E data - is a  task that can be automated with a trained deep neural network. Using the initial dataset Immunocto $V_0$, we trained a model that classifies a 12-channel 64$\times$64 pixels$^2$ image into one of five cell types - CD4$^+$ T cell, CD8$^+$ T cell, B cell, macrophage, and other, non-immune cells. The 12 channels are the RGB channels of the H\&E image, 8 IF channels (Hoechst, CD45, CD3e, CD4, CD8a, CD20, CD163 and CD68) as well as the binary mask from SAM.\\

\noindent Training was done using PyTorch Lightning 2.1.0 and a NVIDIA A10G GPU on Amazon Web Services (AWS). A ResNet50 architecture (MONAI \cite{cardoso2022monai}'s implementation) was trained from scratch using a batch size of 64, a maximum of 25 epochs, a train/validation split of 0.8/0.2, and a starting learning rate of 0.0001 with a cosine annealing scheduler. The Adam optimiser was used with $\epsilon=10^{-7}$, $\beta_1 = 0.9$, and $\beta_2 = 0.999$. The loss function is a weighted cross-entropy function with an $\ell_2$ weight decay of $\lambda=0.0003$ and a label smoothing factor of 0.03. To mitigate class imbalances, the weight for each class is set to the inverse of the number of examples in each class, and then normalised such that all weights sum to 1. \\

\noindent For data augmentation, we use random horizontal and vertical flips with probability 0.4, and a colour augmentation scheme on the H\&E channel to improve robustness to tissue staining variations and different scanners \cite{shen2024deep}. The colour augmentation scheme first extracts hematoxylin and eosin concentrations from H\&E images using the deconvolution approach of Ruifrok and Johnston \cite{ruifrok2001quantification}. Then, stain concentrations are randomly modified using the scheme of Tellez \emph{et al.} \cite{tellez2018whole} (uniform sampling with $\sigma=0.012$). \\

\noindent 5 models were trained using a 0.8/0.2 train/validation split, such that each example is seen exactly once in the validation set across folds. An ensemble model is created where each model is applied and the class probabilities are averaged out over the 5 models. The validation class-wise accuracy is $>$98\% for each immune cell type, demonstrating that immune cell classification is a simple task in the presence of IF channels.

\subsubsection{Step \#5: Automatic generation of a massive database of immune cells}
\label{sect:step5_immunoctov1}
\noindent The model trained on joint H\&E and IF data is applied to the entire cohort of the Orion platform \cite{lin2023high}, which contains 40 colorectal cancer WSIs. Over 54 million objects were detected with SAM on the 40 WSIs; to maximise the quality of automatically generated labels, we only preserve immune cells which have a softmax probability above 95\%. This results in Immunocto $V_1$, the final Immunocto dataset, which contains 2,282,818 immune cells (952,929 CD4$^+$ T cells, 467,395 CD8$^+$ T cells, 480,900 B cells and 381,594 macrophages). Additionally, Immunocto $V_1$ includes 4,565,636 (twice the number of immune cells) other, non-immune cells or objects used to discriminate between immune cells and other entities in the H\&E images. \\

\noindent The data is accessible on the Zenodo platform at (\url{https://zenodo.org/records/11073373}). Section 2 of supplementary material provides additional details on how the database is organised.

\subsection{Quality control of Immunocto}

\noindent The quality of the database is evaluated in three steps: (1) validating SAM's performance as a nuclear segmentation model (section \ref{sect:methods_SAM_validate}), (2) reviewing label quality through via a separability analysis in the IF space (section  \ref{sect:methods_labels_separability}), and (3) a manual expert review of a subset of the dataset (section \ref{sect:methods_patho_review}).

\subsubsection{Validation of SAM for single cell segmentation}
\label{sect:methods_SAM_validate}

\noindent To assess the quality of single cell nuclear contours generated by SAM with no fine-tuning, we compare the nuclei segmentation performance of the original SAM against three fine-tuned SAM variants for nuclear segmentation: MicroSAM \cite{archit2023segment}, CellSAM \cite{israel2024foundation} and CellViT \cite{horst2024cellvit}. Segmentation models are compared on the 6563 lymphocytes annotated in the Lizard \cite{graham2021lizard} testing set, the largest manually labeled pan-cancer dataset for immune cells. The DICE score and recall for lymphocytes are reported in section \ref{sect:results_SAM_validate}.

\subsubsection{Automated label review via clustering}
\label{sect:methods_labels_separability}

\noindent A complete manual quality control on Immunocto $V_1$'s labels (immune cell subtypes) is an impractical task due to the size of the database. Instead, we implemented an automated precision assessment method that calculates the proportion of cells whose subtypes can be definitively identified using their characteristic IF signal patterns.\\

\noindent For a cell of label $S_1$ to be clearly identifiable from a cell with label $S_2$, cell $S_1$ should show high average IF intensity in the relevant channel(s) that identify cell $S_1$ (for instance, CD4 for CD4$^+$ T cells), and low intensity in the relevant channel(s) associated with cell $S_2$ (for instance, CD8 if cell $S_2$ is a CD8$^+$ T cell). The quality of the database is defined by the level of ambiguity in distinguishing subtypes -- poor quality would be seen as overlapping signals, that is, $S_1$ shows strong intensity in both its own and $S_2$'s marker channels. High quality would shown a clear separation between $S_1$ and $S_2$ signal distributions. \\

\noindent We conducted pairwise comparisons of cell intensity distributions across all six possible subtype combinations. To assess the separability between cell types, we fit a linear classifier using scikit-learn's LogisticRegression (version 1.3.2) to establish decision boundaries.  The resulting distributions and classification boundaries are shown in figure \ref{fig:histo_conf}.

\subsubsection{Manual expert review}
\label{sect:methods_patho_review}

\noindent Manual reviewing of a subset of Immunocto $V_1$ was done independently by a board-certified pathologist (KB) and a Senior Lecturer in Oral Pathology (RAE). We randomly selected 1000 cells from Immunocto $V_1$ (200 of each class, including other cells/objects). Manual review was done using Label Studio v1.10.1 on a cell-by-cell level. Raters were provided with the following information: the Immunocto label, $64\times64$ pixels$^2$ H\&E images centered on the cell's centroid along with the registered IF data from the Hoechst, CD45, CD3e, CD4, CD8a, CD20, CD163 and CD68 channels. Larger H\&E images of 256$\times$256 pixels were also included to provide context from  cellular environment around the target cell. We report confusion matrices between each rater and the Immunocto $V_1$ labels, as well as the concordance between the raters via Cohen's Kappa coefficient $\kappa$ (section \ref{sect:results_patho_review}).

\subsection{Evaluation of H\&E classifiers trained on Immunocto}

\noindent To validate Immunocto's effectiveness for clinical applications, we evaluate its ability to train deep learning models to subtype immune cells using H\&E data only (section \ref{sect:methods:immune_cell_subtyping}). We compare these  models with HoverNet, a well established single cell detection model, both on an internal and external databases (section \ref{sect:methods:lympho_identification_internal}).

\subsubsection{Immune cell subtyping}
\label{sect:methods:immune_cell_subtyping}

\noindent An immune cell subtype classifier (CD4$^+$ T cells, CD8$^+$ T cells, CD20$^+$ B cells, CD68$^+$/CD163$^+$ macrophages or other, non-immune cell) was trained on Immunocto $V_1$ using H\&E as well as the binary masks from SAM only. First, training/validation/test was split on a patient-wise level, with 35 used for training, 4 for validation and 1 for testing. Training was done for 50 epochs. The classifier is based on an adapted ResNet50 architecture, allowing us to leverage a pre-trained ResNet50 on ImageNet. Briefly, after applying the first convolutional layer to the H\&E RGB image, the binary mask is encoded with a trainable convolutional layer of the same size (kernel size of 7, stride of 2, padding of 3), and the resulting mask embedding is added to the image embedding. From this point, we refer to this model architecture as the SAM + ConvNet model. Other hyperparameters are the same as the ones listed in section \ref{sect:step4_trainclassifonv0}. \\

\noindent Two approaches are used for classification on H\&E. In the first approach, a classifier was trained using patches of 64$\times$64 pixels$^2$, which contains information about individual cell morphology. When reporting results, this model is presented as morph. (64$\times$64). In the second approach, the classifier was trained with larger patches of 256$\times$256 pixels$^2$, enabling it to integrate information concerning the individual cell as well as its interactions with its immediate surrounding environment. This model is presented as morph. + context (256$\times$256). Results on the hold-out test dataset are shown in table \ref{tab:results_internal}. 




\subsubsection{Internal/External Evaluation}
\label{sect:methods:lympho_identification_internal}

\noindent A comprehensive evaluation of Immunocto involves comparing permutations of architectures/datasets to train a variety of models, and evaluating their performance on various test sets. Models include the SAM + ConvNet approach presented in this manuscript, HoVer-Net \cite{graham2019hover} and the YOLOv10 object detection model \cite{wang2024yolov10}. The SAM + ConvNet models used in this section are the ones focusing on cell morphology - they are trained on patches of 64$\times$64 pixels$^2$ for Immunocto, and patches of 42$\times$42 pixels$^2$ on Lizard to preserve the same physical patch size due to pixel size differences. Training datasets include Immunocto and Lizard. All combinations of architectures and datasets are reported in table \ref{table:res:external_test}. We did not include the HoVer-Net architecture trained on Immunocto due to limited computing resources ($\approx$2000h of training time). Comparison was done on tests sets comprised of: (1) the Immunocto hold-out test set (43706 lymphocytes), (2) the Lizard \cite{graham2021lizard} test set  (6563 manually annotated lymphocytes from multiple cancer sites), and (3) the SegPath \cite{komura2023restaining} test set (28370 lymphocytes from pan-cancer sources). This evaluation also address the generalisation capabilities of model trained with Immunocto (trained on colorectal cancer data)  on other, pan-cancer datasets. \\
 
\noindent Since the HoverNet model trained on Lizard was limited to general lymphocyte detection, we compared models on the task of binary lymphocyte classification (lymphocyte vs. non-lymphocyte). For models trained on Immunocto, a cell is assumed to be a lymphocyte if it is either a CD4$^+$ T cell, CD8$^+$ T cell or a CD20$^+$ B cell. Upon examination of the Lizard and SegPath test sets, we find that only a subset of all lymphocytes in the WSI patches are annotated. In that context, models that find non-annotated lymphocytes and identify them as lymphocytes would register a high false positive rate, which can lead to misleading metrics for comparison. As a consequence, we only report (table \ref{table:res:external_test}) the recall of each model for the general task of lymphocyte detection.\\ 

\noindent For all combinations of trained models and test sets, the test data is resized to match the resolution of the training data, following the pixel sizes of each dataset reported in table \ref{table:res:external_test}. Models trained on YOLOv10 use a confidence level of 0.1, which consistently provided the highest accuracy across models and test sets. \\

\noindent A visual comparison of immune cell classification for a selected ROI in Immmunocto's test set is presented in section \ref{sect:results:lympho_identification_internal} for a subset of the models trained, while a quantitative analysis is reported in section \ref{table:res:external_test}. Another similar visual comparison on a different ROI is provided in section 3 of supplementary material.

\section{Results and discussion}

\subsection{Quality control of Immunocto}

\subsubsection{Validation of SAM for single cell segmentation}
\label{sect:results_SAM_validate}

\begin{figure}[h!]
\centering
{\includegraphics[width=0.8\textwidth]{{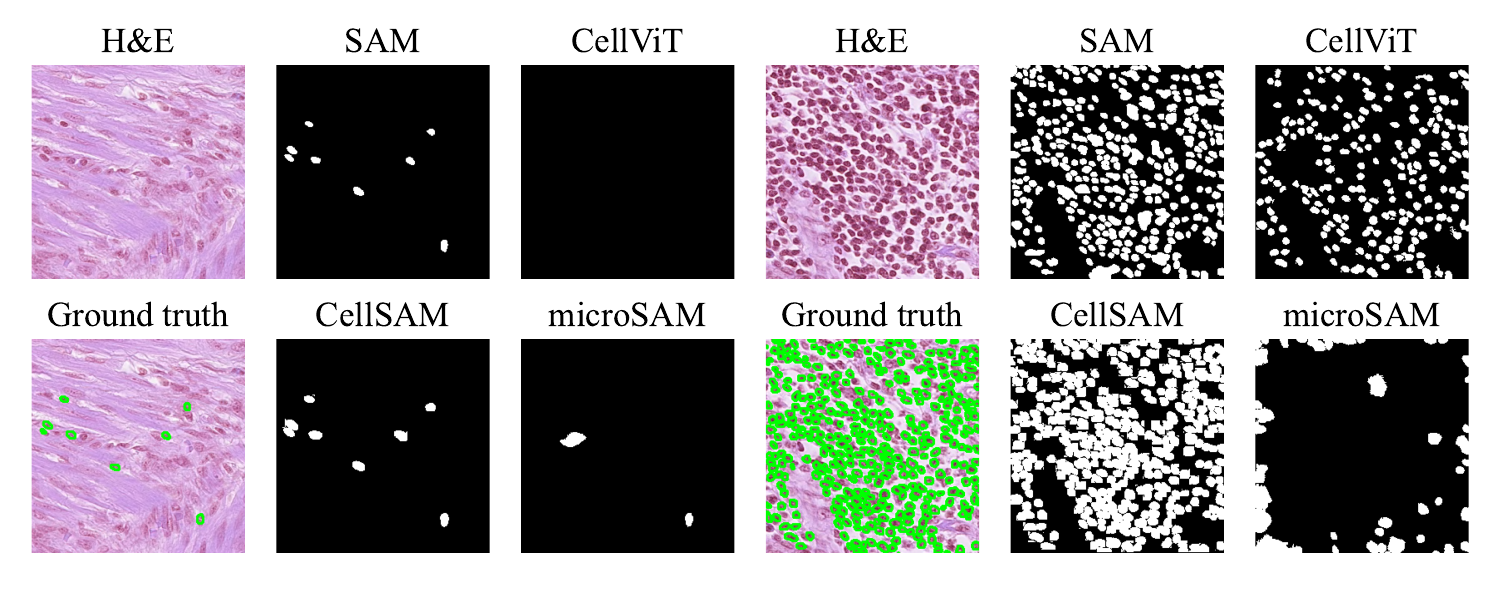}}  
\caption{Examples of the segmentation performance of various segmentation models on the Lizard test set for the identification of lymphocytes only.}
\label{fig:SAM_examples}
}
\end{figure}

\noindent Table \ref{SAM_DICE_Recall} illustrates the DICE coefficient and recall on lymphocyte detection evaluated on the test set of Lizard \cite{graham2021lizard}, for SAM, MicroSAM, CellSAM and CellViT. Figure \ref{fig:SAM_examples} shows the segmentation performance of the four models on two selected ROIs in the Lizard test set.

\begin{table}[h!]
\centering
\begin{tabular}{|c|c|c|c|c|}
\hline
\textbf{Model} & \textbf{SAM} (this work) & \textbf{MicroSAM} \cite{archit2023segment} & \textbf{CellSAM} \cite{israel2024foundation} & \textbf{CellViT} \cite{horst2024cellvit} \\ \hline
\textbf{DICE}  & \textbf{0.762}        & 0.119            & 0.447            & 0.130            \\ \hline
\textbf{Recall} & \textbf{0.981}       & 0.132            & 0.951            & 0.382            \\ \hline
\end{tabular}
\caption{Performance metrics for the segmentation of lymphocytes only evaluated on the test set of Lizard (6563 cells), using various segmentation models.}
\label{SAM_DICE_Recall}
\end{table}

\noindent The above results show that SAM outperforms other models in segmentation and detection tasks, with a high recall indicating that SAM detects the majority of lymphocytes. SAM achieved the highest DICE score among models. It is noteworthy that ground truth contours in Lizard are based on manual annotations; an imperfect DICE score does not necessarily reflect only low segmentation quality, but mainly a disagreement with pathologists, for which there is important inter-observer variability. \\

\noindent CellSAM achieved comparable recall to SAM, but a lower DICE. This is likely due to CellSAM being fine-tuned on the smaller ViT-B architecture of SAM compared to the ViT-H architecture of SAM that we used. Scores of MicroSAM are significantly lower and can be attributed to its lack of fine-tuning on H\&E images and its broader training on a variety of microscopy images. CellViT, in their original inference logs, reported an average recall of 0.58 for detecting inflammatory cells in the PanNuke dataset \cite{gamper2020pannuke}, which is coherent with the limited performance in lymphocyte segmentation seen in table \ref{SAM_DICE_Recall}. Overall, our results suggest that the original SAM with adequate prompting is the most suitable approach to automatically segment single cell nucleus without manual prompting. The high recall obtained on the Lizard test set validates the use of SAM for immune cell detection.

\subsubsection{Automated label review via clustering}
\label{sect:results_labels_separability}

\begin{figure}[h!]
\centering
{\includegraphics[width=0.85\textwidth]{{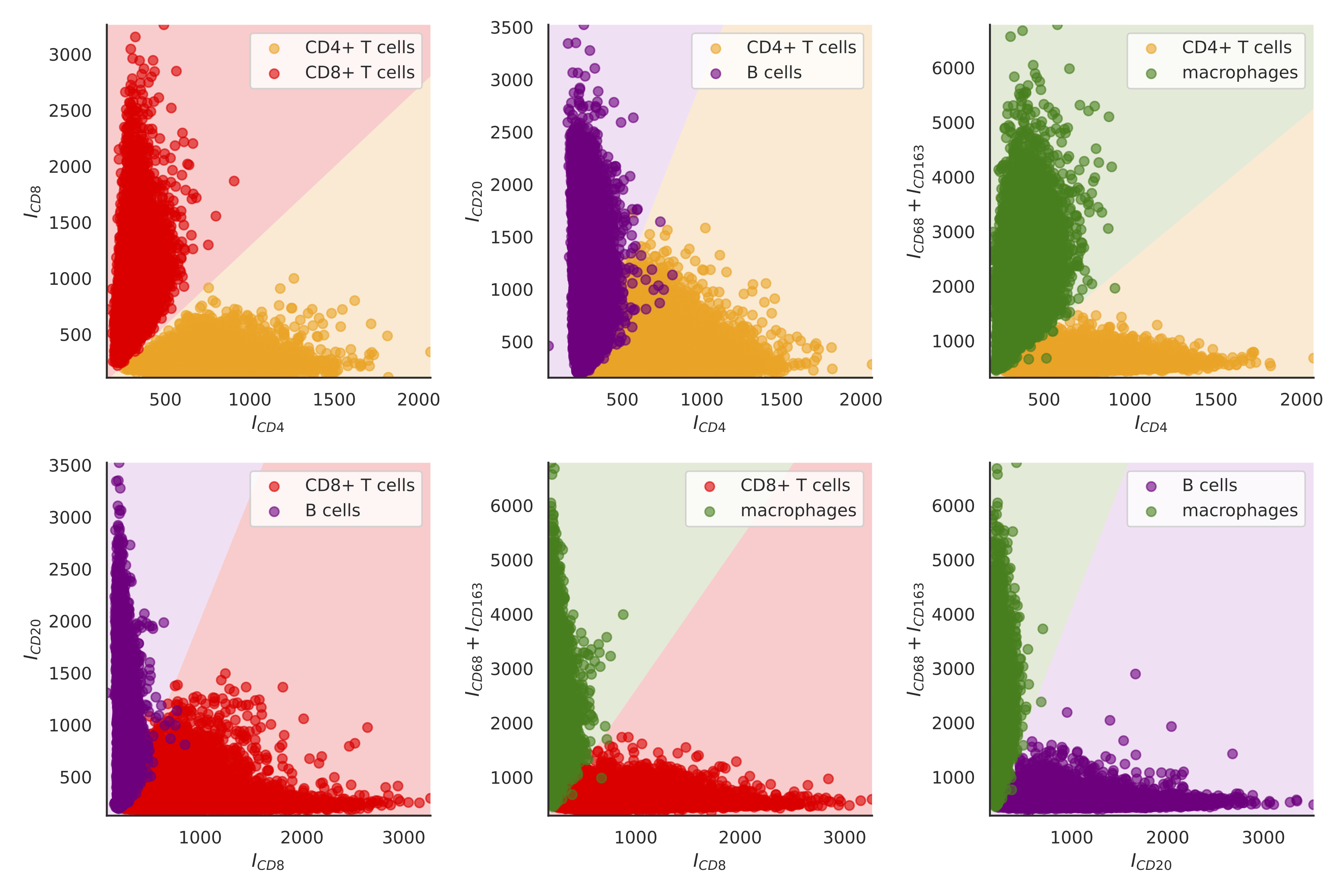}} 
\caption{Differentiation of immune cell subtypes in the Immunocto database based on average IF intensities within cells. For each panel, the compared subtypes are identified in the legend, and the axes $I_{c}$ represent the average intensity of IF channel $c$ inside a cell. Decision boundaries issued from a logistic regression separating the two subtypes are also shown. Only 10\% of the cells (randomly sampled) are shown for each subtype.}
\label{fig:histo_conf}
}
\end{figure}

\noindent In figure \ref{fig:histo_conf}, each panel reports the distribution of Immunocto $V_1$ labels in the space of the two IF channels used to differentiate them. Generally, figure \ref{fig:histo_conf} shows that labels in Immunocto $V_1$ form natural clusters in the space of relevant IF channels.  For all six permutations of immune cell subtypes, we identified the fraction of cells of one subtype $S_1$ that may be ambiguously interpreted as another subtype $S_2$. This is defined as the number of cells from subtype $S_1$ which falls inside the class domain of subtype $S_2$, delimited by the decision boundaries calculated from the logistic regression and shown in figure \ref{fig:histo_conf}. Overall, we find that only 0.51\% of all cells in Immunocto $V_1$ are ambiguously defined. \\

\subsubsection{Manual label review by experts}
\label{sect:results_patho_review}

\begin{figure}[h!]
\centering
{\includegraphics[width=0.85\textwidth]{{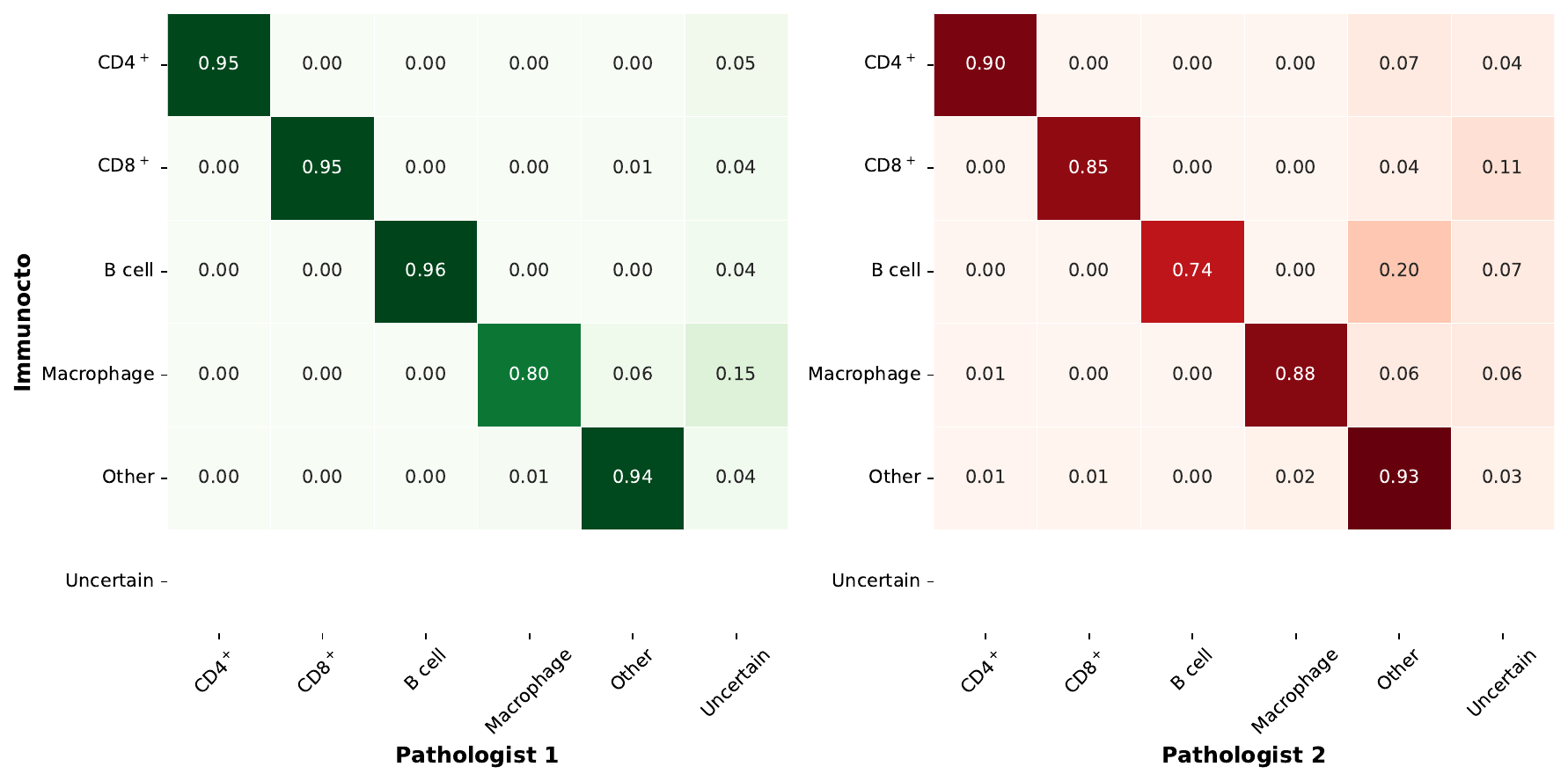}} 
\caption{Confusion matrices showing the agreement between the Immunocto $V_1$ labels and two raters, for each label type. The uncertain label denotes examples where raters could not clearly identify a subtype.}
\label{fig:CM_patho}
}
\end{figure}

\noindent Figure \ref{fig:CM_patho} shows confusion matrices highlighting the concordance between each rater and the Immunocto $V_1$ dataset, for 1000 randomly sampled cells in Immunocto $V_1$. \\

\noindent Averaged over all classes, the agreement between rater 1 and Immunocto $V_1$ is 92\%, while it is 86\% for rater 2. This results in a high average across raters of 89 $\pm$ 4 \%. Overall, the agreement is high for lymphocyte (T cells and B cells) classification, with an average of 89\%. The majority of disagreements are examples where raters were uncertain of the subtype. Such results illustrate the general difficulty in subtyping lymphocytes, and should not be interpreted as a 11\% mislabeling rate by Immunocto. A similar, slightly higher agreement (94 $\pm$ 1\%) and conclusion are found for the label of other cells/objects.\\

\noindent Macrophages have lower agreement between the raters and Immunocto (on average, 84 $\pm$ 6 \%). The morphological spectrum of macrophages is much broader than that of lymphocytes (e.g. depending on phagocytic activity/cytoplasmic digestion). In addition, macrophages intermix with other cell types \cite{hickey2023organization} and usually do not aggregate into clusters, making their identification more difficult. Tangential sectioning in a two-dimensional space adds to the difficulty. Furthermore, we find a substantial agreement between raters, with a Cohen kappa of $\kappa$=0.75, higher than the values ranging from 0.61 to 0.67 observed with the Lizard database for single cell classification \cite{graham2021lizard}.

\subsection{Evaluation of H\&E classifiers trained on Immunocto}

\subsubsection{Immune cell subtyping}

\noindent Table \ref{tab:results_internal} illustrates the performance of the two SAM+ConvNet variants trained on Immunocto (presented in section \ref{sect:methods:immune_cell_subtyping}) to subtype immune cells on H\&E data only, on an internal hold-out patient of the Orion cohort (170,818 objects across the 5 classes). \\

\begin{table}[h!]
\caption{Prediction metrics (precision P, recall R and $F_1$ score) on the Orion hold-out patient per cell type. Results are reported for each model - the morph. 64$\times$64 captures the cell morphology, while morph. + context 256$\times$256 additionally captures the interaction of the cell with its immediate environment. The last two cell types (T cells and lymphocytes) represent aggregated classes.}
\label{tab:results_internal}
\centering
\begin{tabular}{c|cc|cc}
\hline\hline
 & \multicolumn{2}{c|}{morph. (64$\times$64)} & \multicolumn{2}{c}{morph. + context (256$\times$256)} \\ 
Cell Type & P/R & $F_1$ & P/R & $F_1$ \\ 
\hline
B cells & 0.75/0.69 & 0.72 & 0.78/0.80 & \textbf{0.79} \\
CD4$^+$ T cells & 0.60/0.62 & 0.61 & 0.63/0.66 & \textbf{0.65} \\
CD8$^+$ T cells & 0.54/0.59 & 0.56 & 0.57/0.57 & \textbf{0.57} \\
Macrophages & 0.61/0.89 & 0.73 & 0.65/0.94 & \textbf{0.77} \\
Others & 0.99/0.93 & 0.96 & 0.99/0.94 & \textbf{0.96} \\ 
\hline
T cells (CD4 $\&$ CD8) & 0.89/0.95 & 0.92 & 0.91/0.94 & \textbf{0.93} \\
Lymph. (T $\&$ B cells) & 0.92/0.97 & 0.94 & 0.94/0.97 & \textbf{0.95} \\
\hline\hline
\end{tabular}
\end{table}

\noindent We find that classification of immune cells with the larger context window maximises the $F_1$ score for all classification tasks shown in table \ref{tab:results_internal}. The improvement in $F_1$ score is most important for immune cell subtyping (notably B cells, CD4$^+$ T cells and macrophages), while context only provides limited improvements for general lymphocyte detection. Furthermore, the obtained $F_1$ scores per class illustrate that while lymphocyte identification can be reliably achieved using H\&E data only, with an $F_1$ score of 0.95, subtyping lymphocytes and macrophages are challenging tasks, with $F_1$ scores ranging from 0.57 to 0.79.

\subsubsection{Lymphocyte identification - visual evaluation}
\label{sect:results:lympho_identification_internal}

\begin{figure}[h!]
\centering
{\includegraphics[width=0.72\textwidth]{{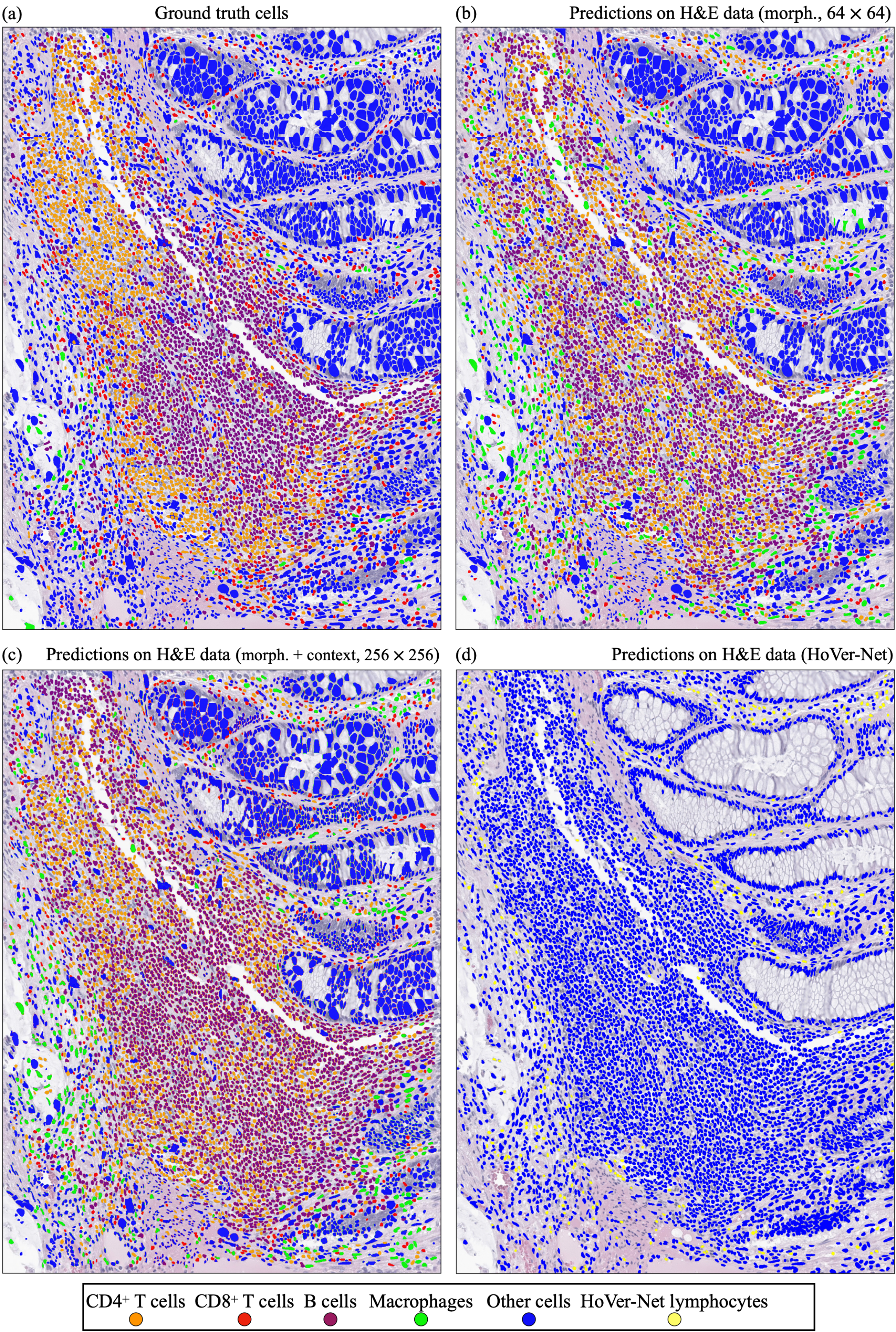}} 
\caption{Overlay of a H\&E ROI containing a high lymphocyte count with predicted cell types from various models. (a) shows the ground truth labels from the Immunocto database; (b) and (c) respectively show the predictions from the morph. 64$\times$64 and morph. + context 256$\times$256 models trained on the remaining Immunocto data. (d) shows the predictions using HoVer-Net.}
\label{fig:side2side}
}
\end{figure}

\noindent Figure \ref{fig:side2side} shows an overlay of a ROI extracted from the Immunocto test set with the segmentation and classification results from three models: the two SAM+ConvNet variants trained on Immunocto, and HoVer-Net trained on Lizard. Another ROI overlay with a lower lymphocyte count is shown in section 3 of supplementary material.

\noindent It is found that classifiers trained on Immunocto outperform HoVer-Net for lymphocyte identification, on two Immunocto test set ROIs. As noted in table \ref{tab:results_internal}, context does not provide significant performance enhancements for the general task of classifying lymphocytes. While the HoVer-Net model has generally appreciable precision, its recall is limited. This can be observed in figure \ref{fig:side2side}d, where most of the lymphocytes could not be detected with HoVer-Net. The difference in performance may be attributed to the large discrepancy in the number of training examples for lymphocytes between the Immunocto and Lizard datasets. In addition, the lymphocyte labels provided by Immunocto are based on IF and may be more robust than the manual ones from Lizard, which have limited inter-observer agreement, as noted in figure C.1 of \cite{graham2021lizard}.

\subsubsection{Lymphocyte identification - Quantitative evaluation}
\label{sect:results:lympho_identification_external}

\noindent The recall on lymphocyte detection for all model/architecture permutations are reported in table \ref{table:res:external_test}.

\begin{table}[h!]
\centering
\begin{tabular}{|c|c|c|c|c|c|c|}
\hline
\multicolumn{2}{|c|}{\textbf{Training Architecture}} & \textbf{HoVer-Net} & \multicolumn{2}{|c|}{\textbf{SAM + ConvNet}} & \multicolumn{2}{|c|}{\textbf{YOLOv10}} \\ \hline
\multicolumn{2}{|c|}{\textbf{Training Dataset}} & Lizard & Lizard & \small Immunocto & Lizard & \small Immunocto \\ \hline
\multirow{3}{*}{\textbf{Test set}} & \small Lizard \tiny $\left(0.5 \frac{\mu \mathrm{m}}{\mathrm{px}} \right)$ & 0.37 & \textbf{0.77} & 0.44 & 0.57 & 0.42 \\ \cline{2-7} 
 & \small SegPath \tiny $\left(0.26 \frac{\mu \mathrm{m}}{\mathrm{px}} \right)$ & 0.10 & 0.45 & 0.20 & 0.04 & \textbf{0.67} \\ \cline{2-7} 
 & \small Immunocto \tiny $\left(0.325 \frac{\mu \mathrm{m}}{\mathrm{px}} \right)$ & 0.08 & 0.23 & \textbf{0.94} & 0.34 & 0.93 \\ \hline
\end{tabular}
\caption{Recall on lymphocyte detections for various testing sets and training sets, with the pixel size of each dataset indicated in parenthesis. For instance, the first element shows the recall on lymphocyte detection evaluated on the test set of Lizard, using the HoVer-Net architecture trained on the Lizard training dataset.}
\label{table:res:external_test}
\end{table}

\noindent The SAM + ConvNet model trained and tested on Immunocto has a recall of 0.94, while the same model trained and tested on Lizard has a lower recall of 0.77. This shows that Immunocto outperforms Lizard in internal dataset testing shown above, and suggests stronger self-consistency for the database. \\

\noindent The YOLOv10 model trained on Immunocto and tested on SegPath has a recall of 0.67, which is higher than the recall of 0.45 of the best performing model trained on Lizard (SAM + ConvNet). This shows that, despite the potential disadvantage of Immunocto not being pan-cancer, it outperforms the pan-cancer Lizard dataset on a third-party dataset. This suggests that the size of Immunocto is a deciding factor in generalisability. Furthermore, lymphocytes are morphologically similar across cancer sites, which may explain the performance. \\

\noindent We observe that inter-database performance is highly variable across models and training datasets. The best example is YOLOv10 trained on Lizard; tested on Lizard, it reaches a state-of-art 0.57 recall, whereas it is 0.04 when tested on SegPath. \\

\noindent While this points towards the need for larger, more diverse datasets to tackle the task of lymphocyte detection, we mainly attribute this variability to the limited quality of lymphocyte labels in Lizard and SegPath. Lizard is heavily based on pathologist review, who have limited concordance amongst experts for lymphocytes. Manual annotations are biased towards well-known lymphocyte morphologies, leaving less conventional morphologies unlabeled \cite{komura2023restaining}. SegPath uses IF, but still produces labels of limited quality due to the use of an automated thresholding procedure alone that misses most lymphocytes. \\ 

\noindent There are some limitations with Immunocto. The database is currently limited to a subset of cells that are important to characterise the TIME. For instance, it does not include neutrophils, plasma cells, eosinophils and fibroblasts. Expanding the database may improve characterisation of the TIME, for instance by helping identifying cancer-associated fibroblasts and tumour-infiltrating neutrophils \cite{binnewies2018understanding}. \\

\noindent Nonetheless, our investigation of external testing allow us to conclude that the improved generalisation performance of models trained on Immunocto can be explained by (1) an adequate methodology to capture a minimally biased set of lymphocytes within the tissue section, and (2) the size of the database. Furthermore, the inclusion of lymphocyte subtypes within the database makes it an essential resource for computational pathology.

\section{Conclusion}

\noindent We propose a methodology to create robust single cell labels from matched H\&E and multiplexed immunofluorescence data, for computational pathology purposes. The methodology is general and can be used to detect other cell types. Using this methodology, we have created Immunocto, a large scale dataset of individual immune cells including CD4$^+$ T cells, CD8$^+$ T cells, CD20$^+$ B cells, and CD68$^+$/CD163$^+$ macrophages. The main advantage of the methodology used to create Immunocto is that it requires minimal manual annotations, as it combines the Segment Anything Model (SAM) \cite{kirillov2023segment} to extract candidate cells and relies on fully registered immunofluorescence (IF) data from the Orion platform \cite{lin2023high} to provide labels for candidate cells. We demonstrate how state-of-the-art results can be obtained for the difficult task of lymphocyte detection using models trained on Immunocto. 

\section{CRediT authorship contribution statement}
\noindent \textbf{Mikael Simard}: Conceptualisation, Methodology, Software, Formal analysis, Data Curation, Writing - Original Draft, Visualisation. \textbf{Zhuoyan Shen}: Conceptualisation, Software, Formal Analysis, Visualisation, Writing - Review \& Editing. \textbf{Konstantin Br\"autigam}: Validation, Writing - Review \& Editing. \textbf{Rasha Abu-Eid}: Validation, Writing - Review \& Editing. \textbf{Maria A. Hawkins}: Writing - Review \& Editing. \textbf{Charles-Antoine Collins-Fekete}: Conceptualisation, Software, Resources, Writing - Review \& Editing, Supervision, Project administration, Funding acquisition.

\section{Declaration of competing interest}
\noindent The authors declare that they have no known competing financial interests or personal relationships that could have appeared to influence the work reported in this paper.

\section{Code and data availability}

\noindent Immunocto is accessible at (\url{https://zenodo.org/records/11073373}). Section 2, supp. material provides extra database details. We provide relevant code at \url{github.com/mikaelsimard5/Immunocto.git}. 

\section{Acknowledgements}
\noindent This project is supported by the UKRI Future Leaders Fellowship, No. MR/T040785/1, the Radiation Research Unit at the Cancer Research UK City of London Centre Award C7893/A28990, as well as the UKRI AI for Health Award EP/Y020030/1. KB was funded by the Swiss National Science Foundation (P500PM\_217647/1). \\
The authors declare no competing interests.


\bibliographystyle{unsrt}

\end{document}